\documentclass[preprint,prc,aps,showpacs]{revtex4}
\usepackage{amsmath,graphicx,epsfig}
\usepackage{bm}
\begin{document}

\title{Two-Photon Interactions with Nuclear Breakup in Relativistic Heavy Ion Collisions} 

\author{Anthony J. Baltz$^1$, Yuri Gorbunov$^2$, Spencer R. Klein$^3$ and Joakim Nystrand$^4$} 

\affiliation{$^1$Brookhaven National Laboratory, Upton, NY, 11973, USA\break
$^2$Physics Department, Creighton University, Omaha, NE, 68178 \break 
$^3$Lawrence Berkeley National Laboratory, Berkeley, CA 94720, USA\break
$^4$Dept. of Physics and Technology, University of Bergen, Bergen, Norway}

\begin{abstract}

Highly charged relativistic heavy ions have high cross-sections for two-photon interactions. 
The photon flux is high enough that two-photon interactions may be 
accompanied by additional photonuclear interactions.  Except for the shared impact parameter,
these interactions are independent.  Additional interactions like mutual Coulomb excitation
are of experimental interest, since the neutrons from the nuclear dissociation
provide a simple, relatively unbiased trigger.

We calculate the cross
sections, rapidity, mass and transverse momentum ($p_T)$ distributions for
exclusive $\gamma\gamma$ production of mesons and lepton pairs, and for
$\gamma\gamma$ reactions accompanied by mutual Coulomb dissociation. 
The cross-sections
for $\gamma\gamma$ interactions accompanied by multiple neutron emission ($XnXn$) and 
single neutron emission ($1n1n$) are
about 1/10 and 1/100 of that for the unaccompanied $\gamma\gamma$ interactions.
We discuss the accuracy with which these cross-sections may be calculated. 
The typical $p_T$ of $\gamma\gamma$
final states is several times smaller than for comparable coherent photonuclear interactions,
so $p_T$ may be an effective tool for separating the two classes of interactions.

\end{abstract}

\pacs{PACS  Numbers: 25.20.-x, 12.40.Vv, 13.60.Le}

\maketitle
%\narrowtext

\section{Introduction}

With their large charges $Z$, relativistic heavy ions carry strong
electromagnetic fields which act as strong sources of nearly-real
photons.  These photons can induce a wide variety of
photonuclear and two-photon physics \cite{reviews}. The photon flux
scales as $Z^2$, so the two-photon cross-section scales as $Z^4$.  
The particle production rates are competitive with
those obtained at $e^+e^-$ colliders.  At the LHC, there is strong interest in
using two-photon physics to search for signs of new physics \cite{LHC}. 

Because the coupling constant $Z\alpha\approx0.6$ ($\alpha\approx
1/137$ is the electromagnetic coupling constant) is large, two-photon interactions
may be accompanied by additional
electromagnetic interactions, such as additional two-photon reactions, or
photonuclear interactions. Photonuclear reactions can produce collective nuclear excitations,
e.g. a Giant Dipole Resonance (GDR), or can involve more energetic
processes, such as pion production. Multiple additional reactions also occur.  For example, one
photon from each nucleus can dissociate the other, in mutual Coulomb dissociation 
(MCD) \cite{mutual,BCW}. Experimentally, neutrons from the nuclear dissociation are detected
in zero degree calorimeters (ZDCs) downstream from the collision points, in
both directions.   These neutrons make a
convenient trigger, separating two-photon and other
ultra-peripheral collisions (UPCs) from backgrounds such as cosmic
rays and beam-gas interactions.  The neutrons also 'tag' events
with a smaller mean impact parameter than exclusive two-photon events.
Neutron tagging has been used to study $e^+e^-$ \cite{STARee,vladimir}
and vector meson photoproduction \cite{STAR,PHENIX,rhotag}.  

Tagging is of special interest for $e^+e^-$ because the stronger fields at small impact parameters,
where non-perturbative or strong-field electrodynamic effects will be strongest. 
Calculations of $e^+e^-$ pair production accompanied by MCD 
are challenging, because the typical impact parameters
are smaller than the electron Compton wavelength, $\lambda_C = 386$ fm
\cite{bauree}.  The cross-section measured by STAR is larger
than lowest-order QED calculation; the difference can be explained 
by higher order corrections \cite{baltzee}.

Here, we calculate the cross sections, rapidity, invariant mass, and
transverse momentum distributions for two-photon production of a
selection of scalar and tensor mesons and $\mu^+\mu^-$ and 
$\tau^+\tau^-$ pairs.   We calculate the cross-sections and 
distributions with and without
nuclear breakup, and explore how nuclear breakup affects the
reactions.  

Because the cross-section for n-photon interactions scales 
as $Z^{2n}$ we focus on heavy nuclei: 
gold-gold collisions at a center of
mass energy of $\sqrt{s_{nn}} = 200$ GeV per nucleon (beam Lorentz boost $\gamma=108$), as seen at the Relativistic Heavy Ion
Collider (RHIC) at Brookhaven National Laboratory, and lead-lead
collisions at an energy of 5.5 TeV per nucleon, as will be produced at
the large hadron collider (LHC) at CERN, where $\gamma=2940$.
We assume that RHIC runs for $10^7$ s/year, at an 
average gold-gold luminosity of
$2\times10^{26}$/cm$^2$/s \cite{usrates}.  The LHC is expected to devote
$\sim$ 1 month, or $10^6$ s/year to lead beams,  at an
average luminosity of $10^{27}$/cm$^2$/s \cite{usrates}.

Section II will present a calculation of the two-photon luminosity.
Section III discusses the effects of multiple interactions, while
Section IV finds the two-photon luminosity under various different
nuclear breakup conditions and discuss the uncertainties in the calculation. 
The final state $p_T$ spectra are discussed in Section V, and Section VI gives 
some conclusions.

\section{Two-Photon Luminosity}

According to the method of equivalent photons, 
the cross section for a two-photon reaction, Fig. 1a, 
factorizes into an elementary cross section for $\gamma\gamma\rightarrow X$ and 
a  $\gamma\gamma$ luminosity ${\cal L}_{\gamma \gamma}$ \cite{brodsky,review1}.  
The cross section to produce a final state with mass $W$ is
\begin{equation}
\sigma(A\!+A\!\rightarrow\!A\!+\!A\!+\!X)=
\int dk_1 dk_2 \frac{n(k_1)}{k_1} \frac{n(k_2)}{k_2} \sigma(\gamma\gamma\rightarrow X(W))
\end{equation}
where $k_1$ and $k_2$ are the two photon energies, and $n(k)$ is the photon flux at energy $k$.

\begin{figure}[bt]
\center{\includegraphics*[width=0.7\linewidth]{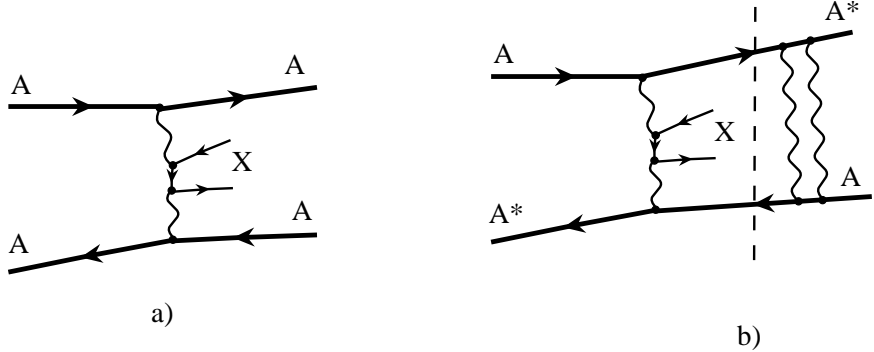}}
\caption[]{The dominant Feynman diagrams for two photon reactions (a)
without and (b) with nuclear excitation.  Cross diagrams and, for (b)
time reversed diagrams are omitted; due to factorization, they
simply add.}
\label{feynman}
\end{figure}

The $\gamma\gamma$ luminosity is given by convolution of the equivalent photon 
spectra from the two nuclei. In impact parameter space, the total number of 
photons from one nucleus is obtained by integrating over all impact parameters larger 
than some minimum, typically given by the nuclear radius. This is similar to 
integrating over all possible momentum transfers, $Q$, from some minimum 
determined by the kinematics up to a maximum given by the nuclear form factor. 
In hadronic collisions, the impact parameter representation provides the best 
way to incorporate effects of strong absorption. Hadronic interactions 
will dominate in collisions where both hadronic and electromagnetic interactions 
are possible. The hadronic interaction must therefore be excluded to obtain the 
effective or usable cross section for two-photon interactions. In impact parameter 
space, this can be accomplished by requiring that the impact parameter be larger 
than the sum of the nuclear radii. The equivalent two-photon luminosity is 
thus \cite{bff,joakim98}:
\begin{equation}
{\frac{d {\cal L}_{\gamma \gamma}}{dWdy}} = 
{\cal L}_{AA}{\frac{W}{2}}\int_{b_1>R_A} d^2b_1
\int_{b_2>R_A} d^2b_2\ n(k_1,b_1) n(k_2,b_2)
\Theta(|\vec{b_1}-\vec{b_2}| -2R_A)
\label{eq:luminosity}
\end{equation}
where ${\cal L}_{AA}$ is the ion-ion luminosity, $n(k,b)$ is the flux of
photons with energy $k$ at impact parameter $b$, and $R_A$ is the Woods-Saxon nuclear radius.

The requirements $b_1>R_A$ and $b_2>R_A$ ensure that the final state is
produced outside the nuclei.  Otherwise, the final state will usually
interact with the nucleus, destroying itself and breaking up the
nucleus.  This requirement may not be strictly necessary for some
final states, such as lepton pairs. Alternately, a smaller radius might be appropriate. However, since the electric fields drop rapidly for $b<R_A$, this is a relatively small correction.  The
$\Theta$ function imposes a requirement that the nuclei not physically
collide; we discuss more detailed hadronic interaction models
below.

The photon flux may be modelled using the Weizs\"acker-Williams
method.  For $\gamma\gg 1$ 
\begin{equation}
n(k,b) = {\frac{d^3N}{dkd^2b}} = {\frac{Z^2\alpha} {\pi^2 kb^2}} x^2 K_1^2(x)
\end{equation}
where $x = bk/\gamma$.  Here, $K_1$ is a
modified Bessel function.   The two photon energies $k_1$ and $k_2$ determine
the center of mass energy $W$ and rapidity $y$:
\begin{equation}
k_{1,2} = \frac{W}{2} e^{\pm y}
\end{equation}
and
\begin{equation}
y = 1/2 \ln(k_1/k_2).
\end{equation}
The maximum effective two-photon energy, $W_{max}$ occurs at $y=0$, when
$k_1=k_2=\gamma/R_A$.  $W_{max}$  is about 6 GeV for gold at RHIC,
and 150 GeV for lead at the LHC.  $W_{max}$ is higher for lighter nuclei and 
protons, but the $\gamma\gamma$ luminosity 
per collision is lower, and multi-photon interactions are very rare.

Equation (\ref{eq:luminosity}) treats the nuclei as hard spheres with radius $R_A$. 
Since there is a finite probability to have hadronic interactions at impact parameters
$b>2R_A$, more accurate hadronic interaction
probabilities can be included by modifying Eq. (\ref{eq:luminosity}), to 
\begin{equation}
{\frac{{\cal L}_{\gamma \gamma}}{dWdy}} = {\cal L}_{AA}{\frac{W}{2}}\int_{b_1>R_A} d^2b_1
\int_{b_2>R_A} d^2b_2 n(k_1,b_1) n(k_2,b_2)
[1-P_H(|\vec{b}_1-\vec{b}_2|)]
\label{eq:luminosity2}
\end{equation}
with the hadronic interaction probability
\begin{equation}
P_H(\vec{b})= 1 - \exp\bigg(-\sigma_{nn}\int d^2\vec{r} 
\ T_A(\vec{r})\  T_A(\vec{r}-\vec{b})\bigg).
\end{equation}
$\sigma_{nn}$ is the total hadronic interaction cross section, 52 mb
at RHIC and 88 mb at the LHC \cite{usrates}.  We use the total cross
sections, since even an elastic nucleon-nucleon interaction will
break up the nucleus.  The nuclear thickness function is the integral of the
nuclear density, $\rho(r)$
\begin{equation}
T_A(\vec{b}) = \int dz \rho(\vec{b},z) dz
\end{equation} 
where $\vec{b}$ is the impact parameter from the center of the nucleus. 
The nuclear density profile $\rho(r=\sqrt{|\vec{b}|^2+z^2})$ of heavy
nuclei is well described with a Woods-Saxon distribution.  We use
parameters determined from electron scattering data (R=6.38 fm for Au and R=6.62 fm for Pb)~\cite{usrates}.

\begin{figure}
\epsfxsize=0.7\textwidth
\centerline{\epsffile{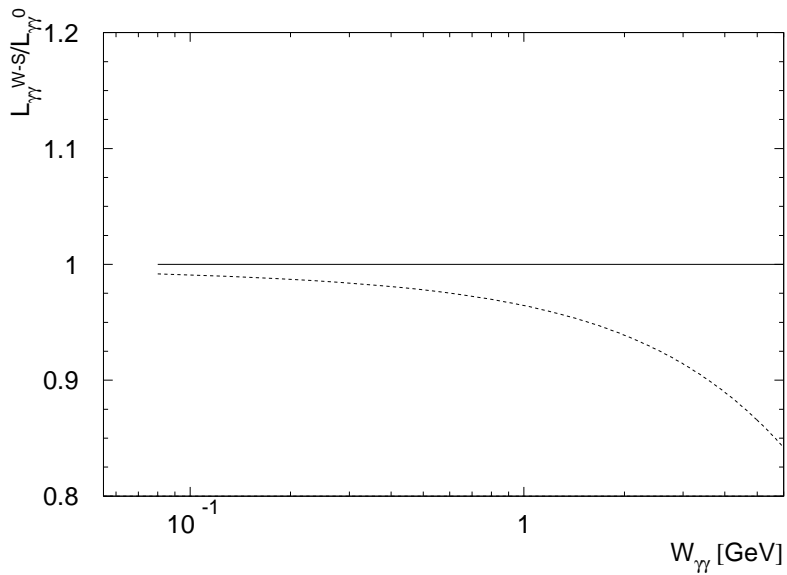}}
\vspace{-0.5cm}
\caption[]{Reduction in two--photon luminosity when the nucleon density is
approximated with a Woods--Saxon distribution (dotted line) as compared
with a flat distribution for hard sphere nuclei, requiring $|\vec{b_1}-\vec{b_2}|>2R_A$ (solid line). 
The calculation is for gold--gold interactions at RHIC.}
\label{wscomp}\end{figure}

As Fig. \ref{wscomp} shows, Eq. (\ref{eq:luminosity2}) gives $\gamma\gamma$ luminosities
about 5\% lower than the hard sphere model for $W=0.1 W_{max}$, falling to
15\% lower for $W=W_{max}$ \cite{joakim98}.  These differences are comparable to those found
elsewhere \cite{roldao}.

Baur and Ferrara-Filho derived Eq. (\ref{eq:luminosity}), and then used a change of
variables and the $\Theta$ function to reduce the dimensions of the
integral \cite{bff}.  Although this approach
speeds the calculation, it works poorly for the
realistic models of nuclear density or when 
additional photon exchange is included.  
Cahn and Jackson used a related approach, calculating the
luminosity analytically without the requirement $b>2R_A$, and then
numerically calculating a correction for the overlap \cite{cj}.  This
approach also cannot accomodate nuclear breakup.

In principle, the meson or lepton pair production cross section depends
on the angle between $\vec{b}_1$ and $\vec{b}_2$, with
different cross sections for parallel and perpendicular photon
polarizations \cite{vidovic}.  However, after integration over $\vec{b}_1$ and
$\vec{b}_2$, the differences are small, and we neglect them.

The cross
section to produce a narrow resonance with spin $J$ and mass $m$ is
\begin{equation} 
\sigma_{\gamma\gamma} = 8\pi^2 (2J+1) \frac{\Gamma_{\gamma\gamma}}{ 2W^2}
\delta(W-m),
\end{equation}
where we neglect the width of the hadronic resonances.  The magnitude of the error
due to this narrow-resonance approximation scales linearly with the width.  
For coherent photonuclear $\rho^0$ production, including 
the width reduces the cross section by about 5\% \cite{usrates}.
A similar correction is expected  for two-photon production; these adjustments should
should scale linearly with the relative width ($\Gamma/M$) of the resonance. 

We also consider the production of continuum lepton pairs.  For
leptons with mass $M$, the cross section is given by
the Breit-Wheeler formula
\begin{equation}
\sigma_{\gamma\gamma} = {4\pi\alpha^2\over W^2}
\bigg[\bigg(2 + {8M^2\over W^2}-{16M^4\over W^4}\bigg)
\ln{W+\sqrt{W^2-4M^2}\over 2M} - \sqrt{1-{4M^2\over W^2}}
\bigg(1+{4M^2\over W^2}\bigg)\bigg].
\end{equation}
This equivalent-photon approach is simpler than a full QED calculation in that it 
neglects the virtuality of the intermediate photon lines in Fig. 1.   For this
reason, we do not consider $e^+e^-$ pairs, where this intermediate state affects
the pair $p_T$ distribution \cite{STARee,bauree}.

\section{Two-Photon Reactions accompanied by Nuclear Breakup}

Because $Z\alpha$ is large, two-photon reactions may be
accompanied by additional photonuclear reactions, as in 
Fig. (1b).  As long as the two meson-producing photons do not excite the nucleus that
emitted them, this is the dominant diagram for producing a meson while exciting both nuclei.
As with photoproduction of vector mesons accompanied by MCD \cite{STAR,rhotag},
the kinematics of photon emission does not favor nuclear excitation.  Then, the individual photon
reactions are independent \cite{gupta} and the processes factorize \cite{factorize}.
Photonuclear breakup can be incorporated in Eq. (\ref{eq:luminosity2}) by including
the photonuclear excitation probability, $P(b)$:
\begin{equation}
{{\cal L}_{\gamma \gamma}\over dWdy} = {\cal L}_{AA}{\frac{W}{2}}\int_{b_1>R_A} d^2b_1
\int_{b_2>R_A} d^2b_2 n(k_1,b_1) n(k_2,b_2)
P(b)[1-P_H(b)],
\label{eq:sigma}
\end{equation}
where $b= |\vec{b}_1-\vec{b}_2|$.
Here, $P(b)$ is the total breakup condition of interest, whether it is one nucleus or two. 
The two-photon luminosity accompanied by single nuclear excitation was calculated in this approach in 
Ref. \cite{hencken}.
We separately consider two cases.  The first, labelled $Xn$ covers all nuclear excitation,
including high-energy excitations which may emit pions in addition to dissociating the nucleus.
The second,  single neutron emission, $1n$, is a subset of $Xn$, in which a single neutron is
emitted.  

The lowest order probability for Coulomb breakup of a specific nucleus is 
\begin{equation}
P^1_{Xn}(b) = \int_{E_{min}}^{E_{max}} dk {d^3n_\gamma \over d^2b dk} 
\sigma_{\gamma A\rightarrow A^*}(k).
\label{eq:lobreakup}
\end{equation}
where $k$ is the photon energy and $d^3n_\gamma/d^2b dk$ the photon
density from Eq. (3).  $\sigma_{\gamma A\rightarrow A^*}(k)$, 
the excitation cross section, is 
determined by data collected over a wide range of energies \cite{mutualdissoc}; 
$E_{min}$ is the minimum energy for this excitation, while
$E_{max} = \gamma\hbar c/b$ is the maximum photon energy for which there
is significant flux.  
The superscript $1$ shows that this is the lowest order probability.
A similar calculation was performed in Ref.  \cite{mut2}. 

More precisely, $P^1_{Xn} (b)$ is the mean number of
excitations; the probability of having exactly N excitations follows a
Poisson distribution, so the probability for at least one Coulomb excitation is
\begin{equation}
P_{Xn}(b)=1-\exp{(-P^1_{Xn}(b))}.
\end{equation}

Single neutron emission usually occurs when the nucleus is excited
into a giant dipole resonance (GDR), where the protons and neutrons
oscillate against each other \cite{goldhaber}.  Photons with energies
from 8-24 MeV can excite a GDR resonance in heavy nuclei.  
To observe the $1n$ breakup, the
GDR excitation must not be accompanied by any higher-energy reactions:
\begin{equation}
P_{1n}(b)=P^1_{1n}(b) \exp{(-P^1_{Xn}(b))}.
\end{equation}
The lowest order probability $P^1_{1n}(b)$ is determined as in
Eq. (\ref{eq:lobreakup}), except that the photon energy integration is
truncated at the maximum GDR energy, 24 MeV for
both nuclei.   Data on photoemission of
single neutrons is used, avoiding uncertainties in the GDR
branching ratios \cite{GDRreview}. 

In MCD, the two nuclear breakups occur
independently \cite{BCW,mutualdissoc,factorize}, so the probability for
MCD is 
\begin{equation}
P_{XnXn}(b)=(P_{Xn}(b))^2
\end{equation}
and
\begin{equation}
P_{1n1n}(b)=(P_{1n}(b))^2. 
\end{equation}
This independence has been verified by comparing the cross-sections
for single nuclear breakup, both 1n and Xn, and MCD \cite{chiu}.  
It also appears to hold for $\rho^0$
photoproduction with accompanying MCD \cite{STAR}. 

\section{Cross Sections, Rapidity and Mass Distributions}

Table I shows the two-photon widths and cross sections for the different tags at RHIC and the LHC, for several 
meson resonances and for $\mu^+\mu^-$ and $\tau^+\tau^-$, from Eq.~(\ref{eq:sigma}).  For a `standard' RHIC year,
the 550 $\mu$b cross-section
for the $f_2(1270)$ leads to 1,100,000 events/year.  At the LHC,
a $1\mu$b cross section corresponds to 1,000 events/year. 

\begin{table}[bt]
\caption{Spin/parity, two-photon widths $\Gamma_{\gamma \gamma}$ \cite{pdg2008}, and cross sections for 
two-photon production of various final states with and without nuclear breakup.}
\center{\begin{tabular}{lccrrr}
\hline
 \hskip 0  in  Meson \hskip .5 in
&\hskip .5  in $J^{PC}$ \hskip .5 in
&\hskip .5  in $\Gamma_{\gamma\gamma}$ \hskip .5 in
&\hskip .5 in overall \hskip .5 in 
&\hskip .5 in XnXn\hskip .5 in  
&\hskip .5 in 1n1n\hskip .5 in
\\
 	& & (keV) & $\sigma$ [mb]  & $\sigma$ [$\mu$b]  & $\sigma$ [$\mu$b]  \\ 
\hline
\multicolumn{6}{c}{Gold beams at RHIC 
} \\ \hline
$\eta$ 	         &$0^{-+}$ & 0.510$\pm$0.026     & 1.05	   & 51.7   & 4.9	\\
$\eta'$	         &$0^{-+}$ & 4.30$\pm$0.15       & 0.72	   & 49.7   & 4.6	\\
$f_2(1270)$   	 &$2^{++}$ & 2.60$\pm$0.24       & 0.55    & 44.9   & 4.1	\\
$f_{2}'(1525)$ 	 &$2^{++}$ & 0.081$\pm$0.009     & 0.0069  & 0.63   & 0.057   \\
$\eta_c$      	 &$0^{-+}$ & 7.2$\pm$0.7$\pm$2.0 & 0.0029  & 0.40   & 0.034 	\\
$\mu^+\mu^-$     &         &                     & 142     & 5,628  & 537 \\
$\tau^+\tau^-$	 &	   &		         & 0.00079 & 0.14   & 0.011 \\  
\hline
\multicolumn{6}{c}{Lead beams at LHC 
} \\ \hline
$\eta$ 	         &$0^{-+}$ & 0.510$\pm$0.026 	        & 18.8    &  337   & 21.6	\\
$\eta'$	         &$0^{-+}$ & 4.30$\pm$0.15	 	& 21.9    &  469   & 30.3	\\
$f_2(1270)$	 &$2^{++}$ & 2.60$\pm$0.24	 	& 23.4    &  562   & 35.7	\\
$f_{2}'(1525)$   &$2^{++}$ & 0.081$\pm$0.009     	& 0.38    &  9.7   & 0.62 \\
$\eta_c$	 &$0^{-+}$ & 7.2$\pm$0.7$\pm$2.0 	& 0.57    & 19.3   & 1.2	\\
$\mu^+\mu^-$     &         &                     	& 2,017    & 33,084 & 2,128 \\
$\tau^+\tau^-$	 &	   &		         	& 0.55	  & 22.8   & 1.4  \\ 
\hline
\end{tabular}}
\label{sigmaLHC}
\end{table}

Untagged meson production rates have been calculated by many
authors \cite{vidovicgg,joakim98,reviews,chikin,roldao2}.  Unfortunately, some
authors used different ion species and Lorentz boosts. Several earlier
papers considered uranium beams at RHIC with $\gamma=100$.  These papers found
$\eta'$ production cross-sections of 2.9 mb \cite{baurFF} to 3.6 mb \cite{chikin}. For lead at the LHC, with $\gamma=4000$
(rather than the planned $\gamma=2940$ which is used here), they find $\eta'$ cross-sections of 30 and 32 mb
respectively, compared to the 22 mb found here.   Roldao and Natale \cite{roldao2} consider uranium beams at 
RHIC, and lead beams with $\gamma=3360$ at the LHC, using the Cahn-Jackson approach \cite{cj}; our RHIC results are
about 40\% lower than theirs; the LHC results are quite comparable, except that our $\eta_c$ cross-section is 
6 times larger than theirs.  It is difficult to understand the large difference for the $\eta_c$.
The agreement between these calculations is generally good, but detailed comparisons are not possible.

Newer calculations have used the same species and boosts.  Our cross-sections are
0 to 20\% higher than those in Ref. \cite{Bertulani}; this work seems to use $\gamma=100$ for RHIC.
The exception is $\eta_c$ production at RHIC, where the current cross-section is 6 times higher.  
The difference may lie in how the papers handle the non-collision requirement; Bertulani and Navarra
apply a form factor to the nuclei instead of applying an explicit requirement that $b>2R$; this can 
lead to substantially different results near the kinematic limit.  

These rates are also very similar to the $\eta_c$ results reported in Ref. \cite{vidovicgg}.
Our rates are  10-25\% higher than those in a very similar calculation \cite{joakim98}; the
difference stems from slightly different $R_A$ (Ref. \cite{joakim98} used an A-dependent 
parameterization of the nuclear radius, rather than electron scattering data \cite{usrates}) 
and older values for the two-photon widths, $\Gamma_{\gamma \gamma}$. 

The RHIC $\mu^+\mu^-$ and $\tau^+\tau^-$ rates are about 20\% and 33\% respectively lower than
in Ref. \cite{vidovicgg}.  The $\tau$ pairs are close to threshold, so are very sensitive to the 
detailed of the nuclear surface calculation.  For the $\mu$ pairs, the difference may be 
because Ref. \cite{vidovicgg} integrates over rapidity rather than the impact
parameter for the second photons; the effect on the nuclear and final-state overlap requirements are 
not transparent.  The RHIC and LHC $\mu^+\mu^-$ rates are also about 37\% and
22\% respectively lower than approximate Born calculations of
Hencken, Kuraev and Serbo \cite{hks} at roughly the same energies ($\gamma$ of
108 and 3000).  One of us has also performed Born calculations by numerical
integration \cite{ajb09},
and found results consistent with Hencken, Kuraev and Serbo when adjusted for
the slightly lower energies of the numerical calculations ($\gamma$ of 100 and
2760).  
These Born calculations have no cutoff in integration over
$b_1$ and $b_2$; point charges are assumed for the ions and a correction is
made by applying a form factor for the virtual photons emanating from each ion.

\begin{figure}
\center{\includegraphics*[width=0.6\linewidth]{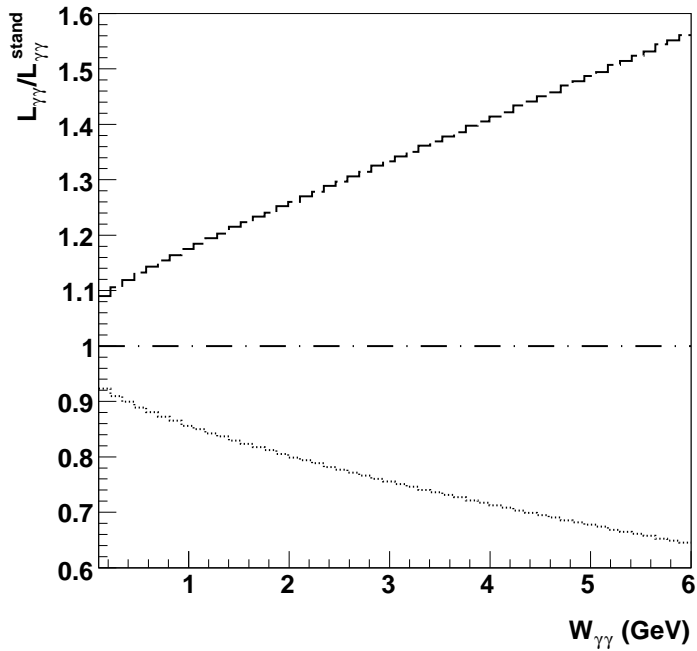}}
\caption[]{ Two--photon luminosity ${\cal L}_{\gamma\gamma}$ as a function of  $W$ for two different nuclear radii (+/-10\% change) for gold beams at RHIC.}
\label{fig:ratio}
\end{figure}

Notable differences between the various calculations are in 
the treatment of the possibility of accompanying hadronic interactions and in
how the virtual photons are cut off, either by integration over $b_1$ and $b_2$ or by a 
form factor.   Here, the ultimate accuracy depends on a 
knowledge of the nuclear matter distribution; although there are accurate measurements of the 
proton distributions in heavy nuclei. Data on neutron radii is sparse, but it appears that neutrons are
less confined than protons in heavy nuclei.
For lead the neutron radius has been estimated 
to be 0.17 fm larger \cite{pbarp}.  Figure \ref{fig:ratio} shows how ${\cal L}_{\gamma\gamma}$ varies with $W$ 
for a +/-10\% change in nuclear radius, for gold beams at RHIC.  A $\pm$ 10\% change in radius leads 
to a $\pm$ 10-30\% change in luminosity.  Uncertainty in the nuclear radius leads to an irreducible 
systematic uncertainty in $\gamma\gamma$ luminosity.

A similar limitation holds for $\gamma\gamma$ interactions
with protons, where the choice of
proton form factor or radius can substantially affect the  photon flux \cite{kn2004}.  Efforts to use lepton pair production to measure the luminosity at the 
LHC \cite{CDFee} have to take into consideration these radius and form factor
uncertainties as well as the possibility of higher order Coulomb corrections
arising from the large Z heavy ions \cite{LHC}.

The $XnXn$ and $1n1n$ cross-sections in Table I are new.  They are about 1/10 and 1/100 of the
untagged cross sections, respectively, at RHIC; at the LHC the reduction is a bit larger.  
These are small fractions, but $XnXn$ reactions may still be of value
because of the triggering advantages that they provide. 

The cross sections and corresponding production rates for single mesons are high. 
For example, one can expect to produce $\approx 10^6$ and $\approx 23 \cdot 10^6$ 
$f_2(1270)$ mesons in a standard RHIC and LHC year, respectively. If one requires mutual 
Coulomb breakup (XnXn) the rates are reduced to $\approx 90,000$ and $\approx 560,000$. 
The $\eta_c$ might be hard to detect at RHIC, but the rates should be sufficient at the 
LHC, where around $600,000$ $\eta_c$ mesons should be produced in one year. Of these, 
$\approx 20,000$ should remain if one requires XnXn breakup. Although these rates are high, 
it should be noted that the cross section for coherent vector meson production with similar 
event topology are about two order of magnitudes larger for mesons of similar 
mass \cite{usrates,rhotag}. A way to separate the two reactions may be through the 
different $p_T$-spectra, as will be discussed in the next section. 

\begin{figure}
\begin{picture}(400,340)

\put(-30,180){\includegraphics{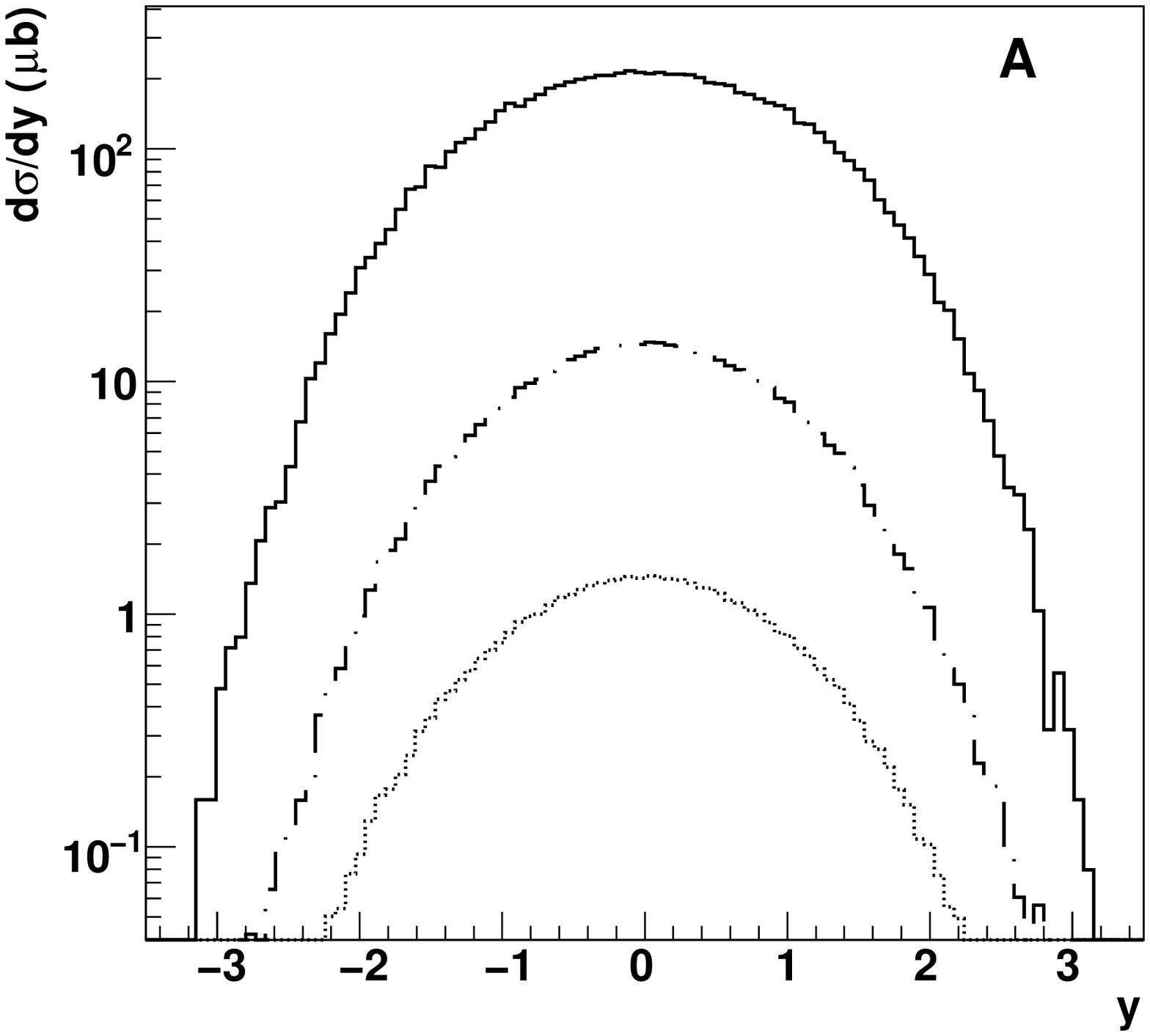}}
\put(200,180){\includegraphics{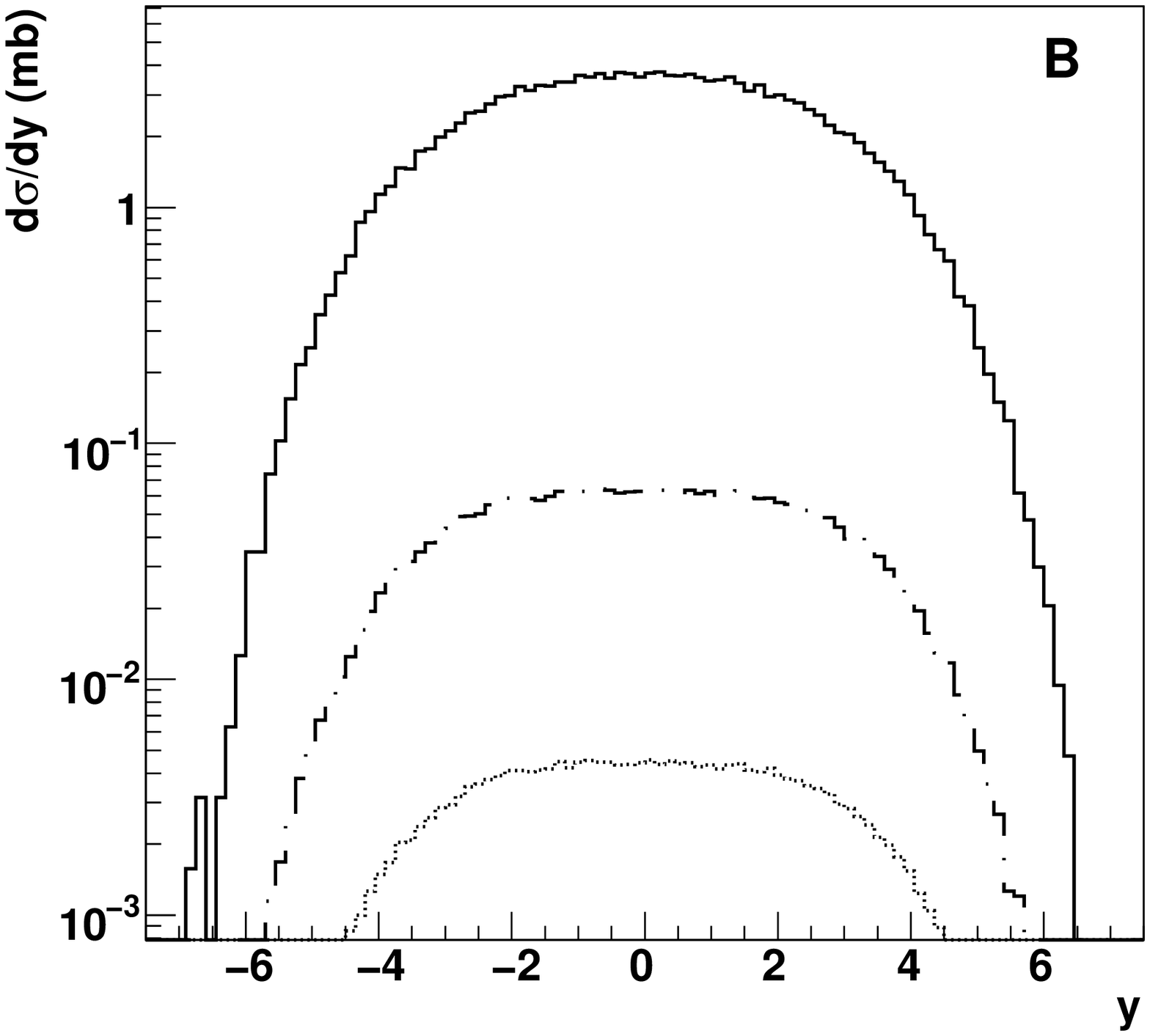}}
\put(-30,0){\includegraphics{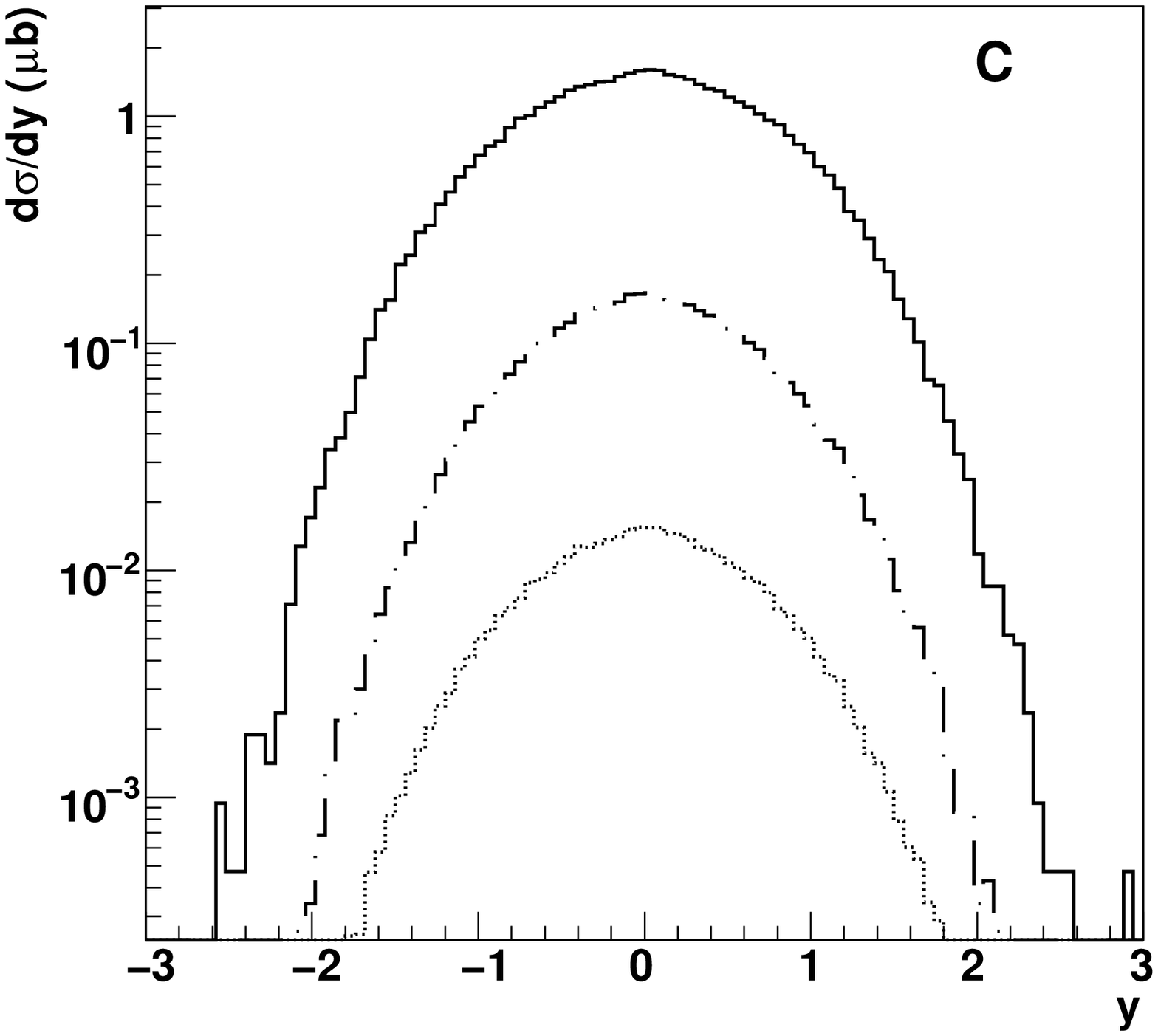}}
\put(200,0){\includegraphics{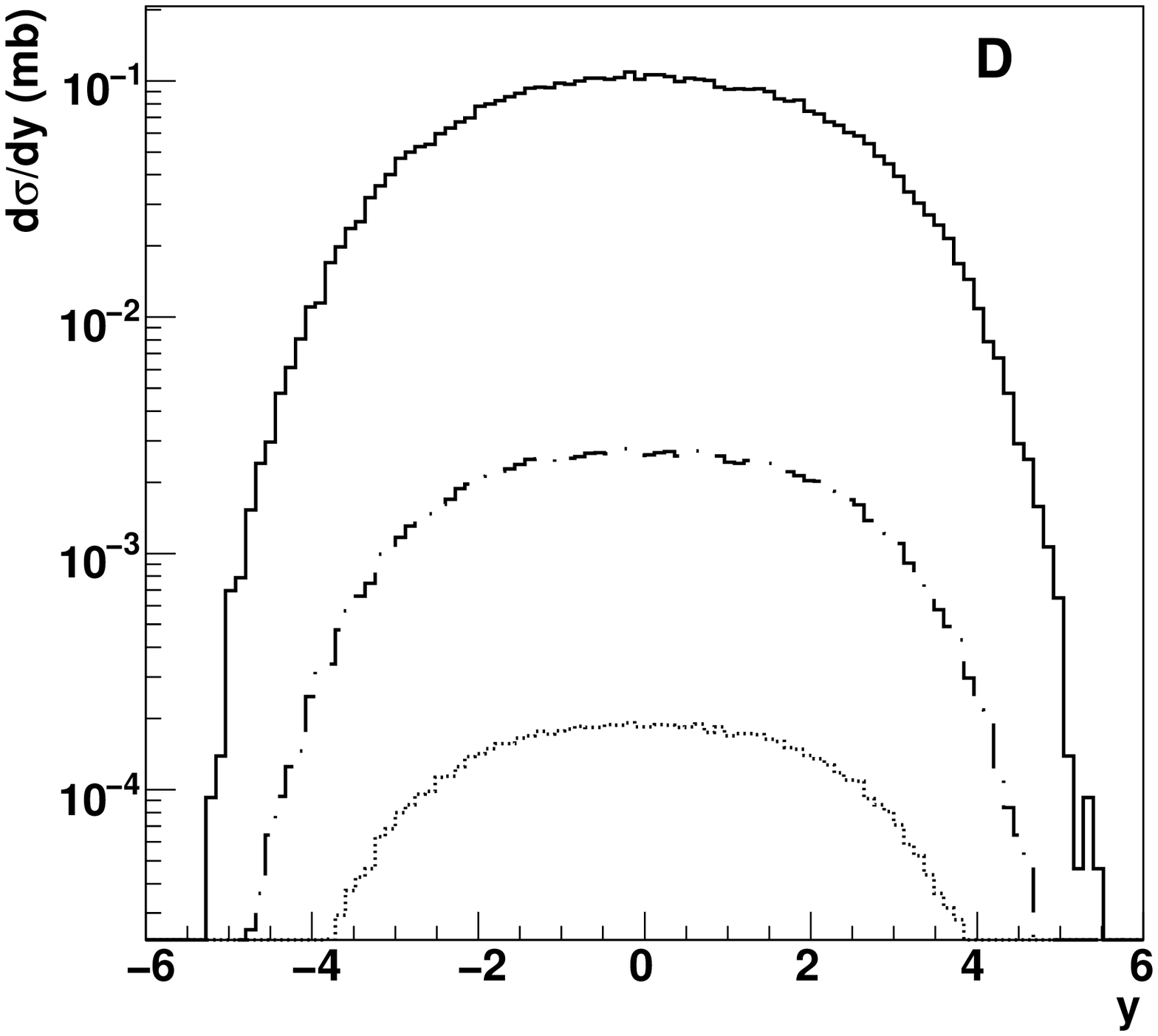}}
\end{picture}

\caption[]{The rapidity spectrum $d\sigma/dy$ for $f_2(1270)$ with
(a) gold beams at RHIC and (b) lead beams at the LHC,
and for the $\eta_c$ with (c) gold at RHIC and (d) lead at the
LHC. Three curves are shown in each panel: the total two-photon cross
section (solid curve), and with $XnXn$ (dashed curve) and $1n1n$
(dotted curve) excitation.}
\label{dndy}
\end{figure}

Figure \ref{dndy} shows the rapidity distribution $d\sigma/dy$ for
$f_2(1270)$ and $\eta_c$ production at RHIC and the LHC.  Spectra for
$XnXn$ and $1n1n$ breakup are shown, along with the untagged
$d\sigma/dy$.  The distributions with breakup are slightly narrower than those
without.  This is because MCD selects events with smaller
impact parameters.  This increases the fraction of events with hard
photons, eliminating interactions at large impact parameter, where a
low energy photon far from it's source nucleus interacts with a higher
energy photon near it's source; the large energy disparity corresponds
to a large rapidity.

\begin{figure}

\begin{picture}(400,340)
\put(-30,180){\includegraphics{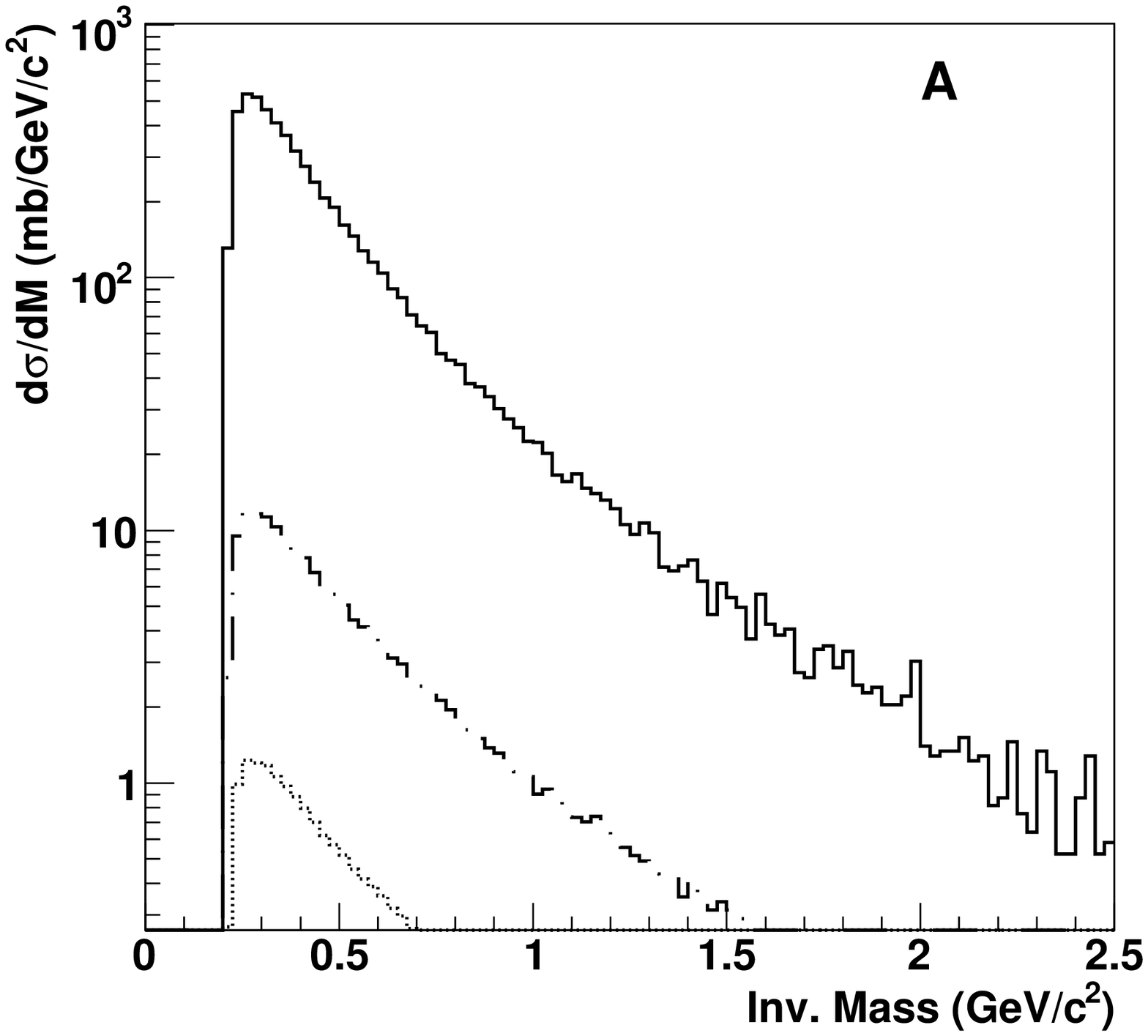}}
\put(200,180){\includegraphics{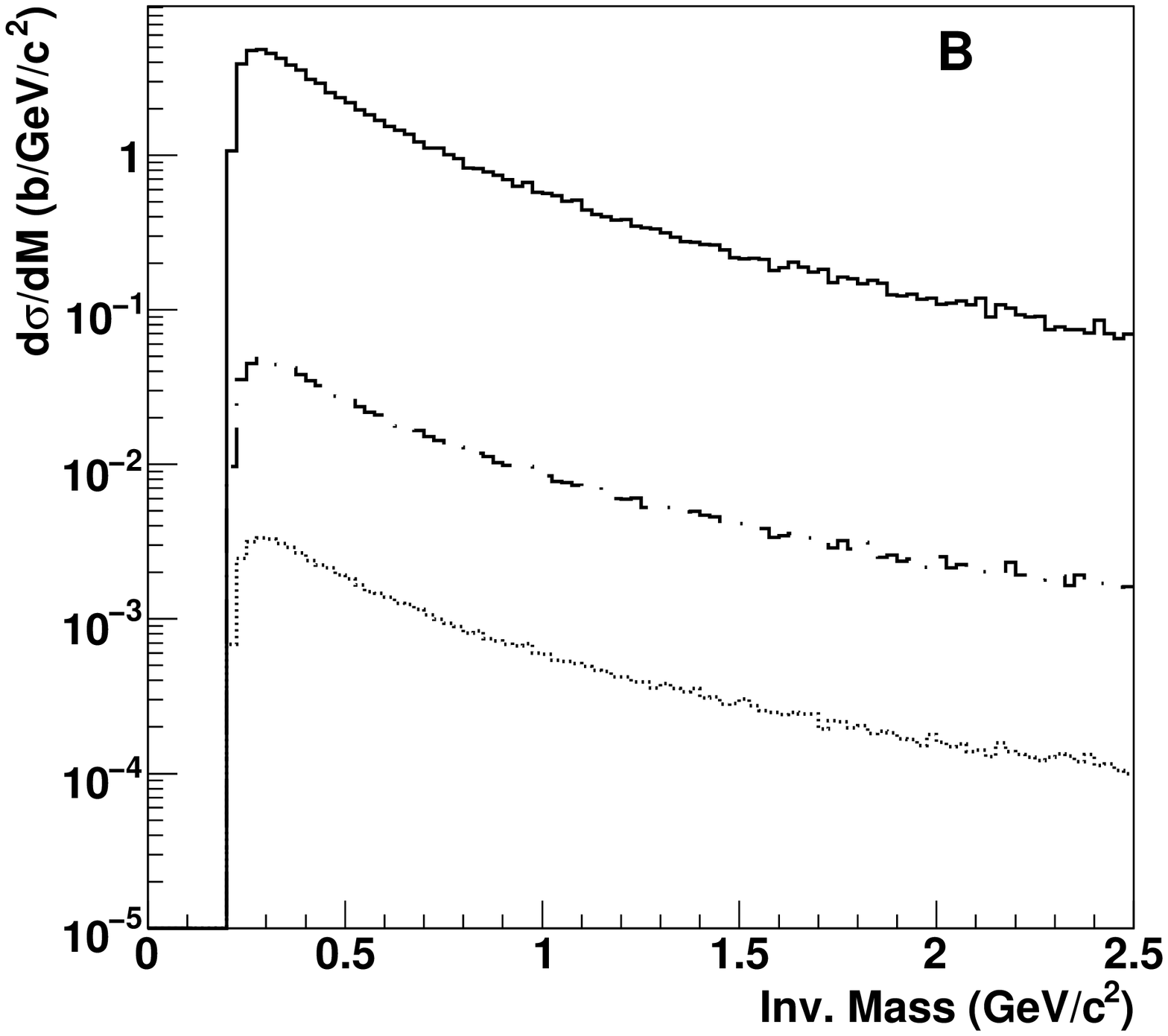}}
\put(-30,0){\includegraphics{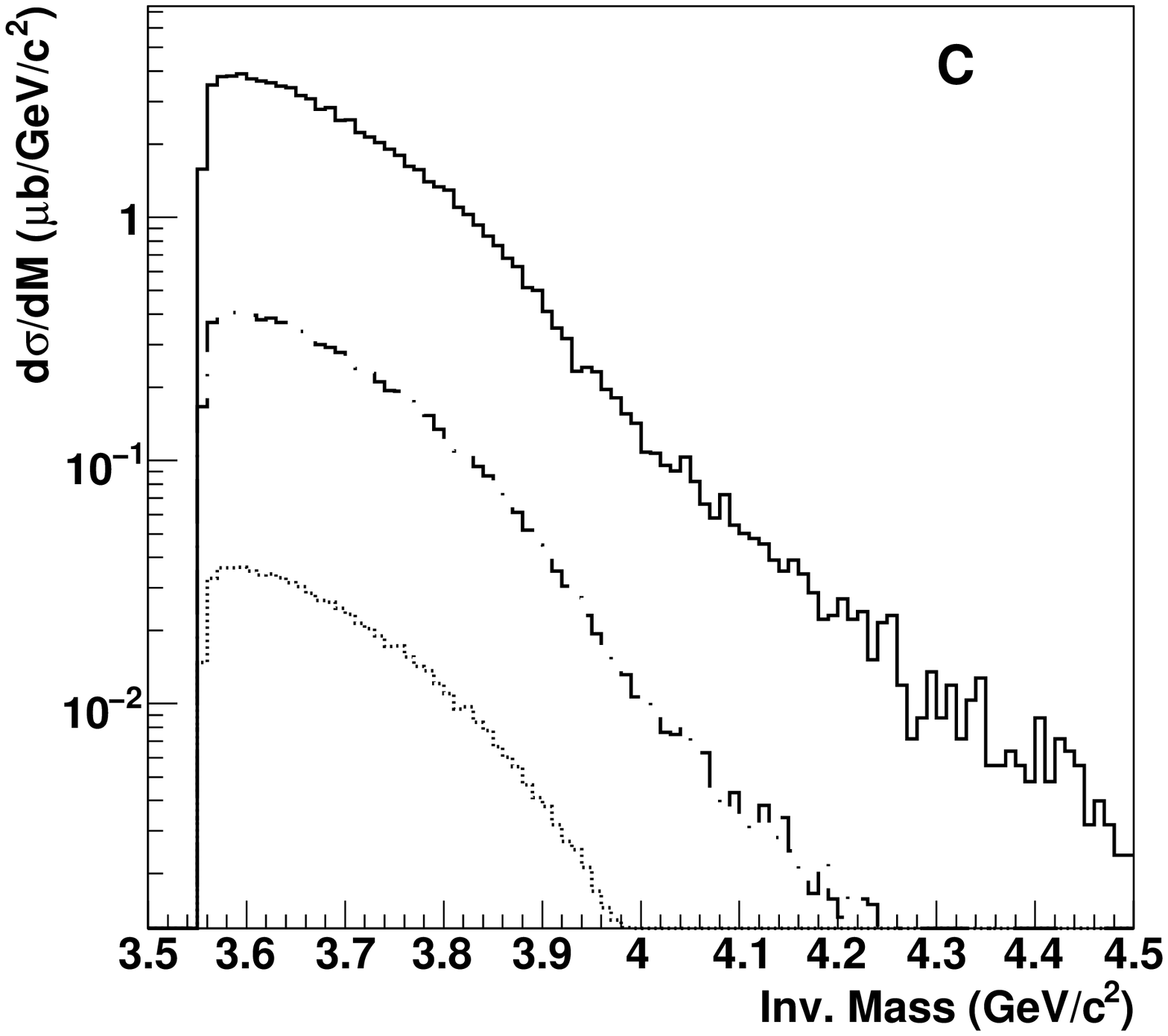}}
\put(200,0){\includegraphics{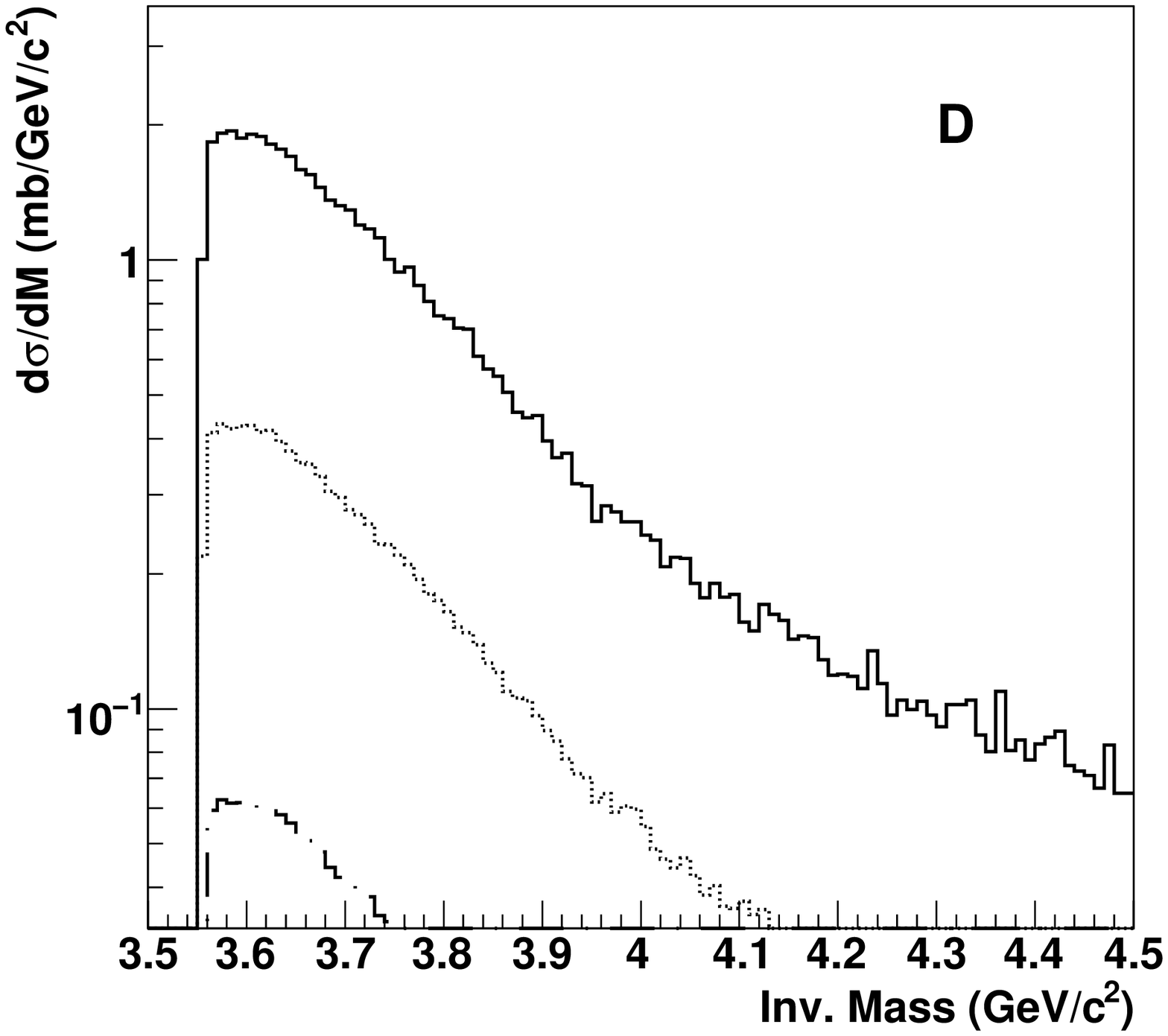}}
\end{picture}
\caption[]{The invariant mass spectrum $d\sigma/dM$ for $\mu^+\mu^-$ pairs with 
(a) gold beams at RHIC and (b) lead beams at
the LHC, and for the $\tau^+\tau^-$ with (c) gold
at RHIC and (d) lead at the LHC.  Three curves are shown in each
panel: the total two-photon cross section (solid curve), and with
$XnXn$ (dashed curve) and $1n1n$ (dotted curve) excitation.}
\label{pairmass}
\end{figure}

Figure \ref{pairmass} compares the $\mu^+\mu^-$ and $\tau^+\tau^-$ pair mass distributions at 
RHIC and the LHC.  Subject to the
aforementioned caveats about beam species and energies, the untagged
spectra are similar to those found by other authors \cite{reviews,vidovicgg,leptonrefs}.
The solid line shows the untagged spectra, while the dashed and
dotted lines are for $XnXn$ and $1n1n$, respectively.   The mass spectra
with breakup are harder than those without. 

\section{Transverse Momentum Spectra}

Ultra-peripheral interactions (UPCs) are fully coherent with both nuclei, and
so have a very small final state $p_T$; this characteristic 
is important for separating UPCs from background events.  Here we consider a more complicated problem,
that of separating fully reconstructed $\gamma\gamma$ interactions from coherent photonuclear interactions.
The final state meson $p_T$ is the vector sum of the two photon perpendicular
momenta, $k_\perp$.  We assume that, integrated over all of
transverse space this angular distribution is flat, and add the two-photon $k_\perp$ in quadrature.

At a distance $b$ from the center of the emitting nucleus the photon flux is~\cite{vidovic}
\begin{equation}
{d^3n(b,k) \over dkd^2b}\! =\! {Z^2\alpha\over\pi^2 k}
\bigg|\!\int_0^\infty\!dk_\perp 
{F(k^2_\perp + k^2/\gamma^2)k_\perp^2
\over  k_\perp^2 + k^2/\gamma^2 }
J_1(bk_\perp)\bigg|^2
\label{photonflux}
\end{equation}
where $J_1$ is a  Bessel function.

For heavy nuclei, the nuclear charge form factor $F$ can be
analytically modelled by the convolution of a hard sphere with a
Yukawa potential of range $a= 0.7$ fm \cite{usrates}:
\begin{equation}
F(q) = {4\pi\rho_0 \over Aq^3} \big[ \sin(qR_{WS}) - qR_{WS}\cos(qR_{WS})\big]
\big({1\over 1+a^2q^2}\big)
\end{equation}
Here, $\rho^0$ is the density and $R_{WS}$ is the Woods-Saxon radius of the nucleus. This gives an excellent 
approximation to the Woods-Saxon distribution. The radii of the heavy ions are well measured however 
recently published results indicate 
that  the neutron and proton distributions differ in the nuclei which may lead 
to a larger radii of the matter distribution and essentially limit the accuracy 
of the photon flux  determination~\cite{pbarp}.   

The $k_\perp$ spectra of the virtual photon fields are
complicated because $k_\perp$ and transverse position are conjugate
variables, and cannot both be defined at the same time.  This
complicates any determination of the $k_\perp$ spectrum with
constraints on transverse position, such as those imposed by
photoexcitation reactions.   We avoid this problem by
selecting the photon energies, and then determining the transverse
momenta solely by using the energies, using
Eq. (\ref{photonflux}).  The photon $k_\perp$ spectrum for fixed $k$ is given by~\cite{usinterf}, 
\begin{equation}
{dN \over dk_\perp} = {2 Z^2\alpha F^2(k_\perp^2+k^2/\gamma^2) k_\perp^3 
\over \pi [k_\perp^2+k^2/\gamma^2]^2}.
\label{eq:pperp}
\end{equation}

This is the same approach used to calculate $p_T$ spectra for vector meson photoproduction \cite{usinterf}.
Hencken and collaborators used Eq. (\ref{eq:pperp}), with a point-particle form factor 
and a cutoff $p_T<1/R_A$ \cite{hencken}.  Other groups (including some of the
present authorship) used this equation with a Gaussian form factor
\cite{starnote}. 

The photon $k_\perp$ distribution is asymmetric, with a 
maximum at $k_\perp =\sqrt{3}k/\gamma$ and large tail at high $k_\perp$.  The
distribution has two different momentum scales, $\hbar/R_A$ and
$k/\gamma$.  For $\hbar/a > k_\perp\gg\hbar/R_A$ and $k_\perp\gg k/\gamma$,
the tail falls as $1/k_\perp^{5}$, regulated by the form factor.

The final state $p_T$ is the vector sum of the two photon $k_\perp$.  We
assume that the angle between the two photons is completely random,
neglecting the possibly different cross sections for photons with
parallel and perpendicular spin.  

\begin{figure}
\center{\includegraphics*[width=0.75\linewidth]{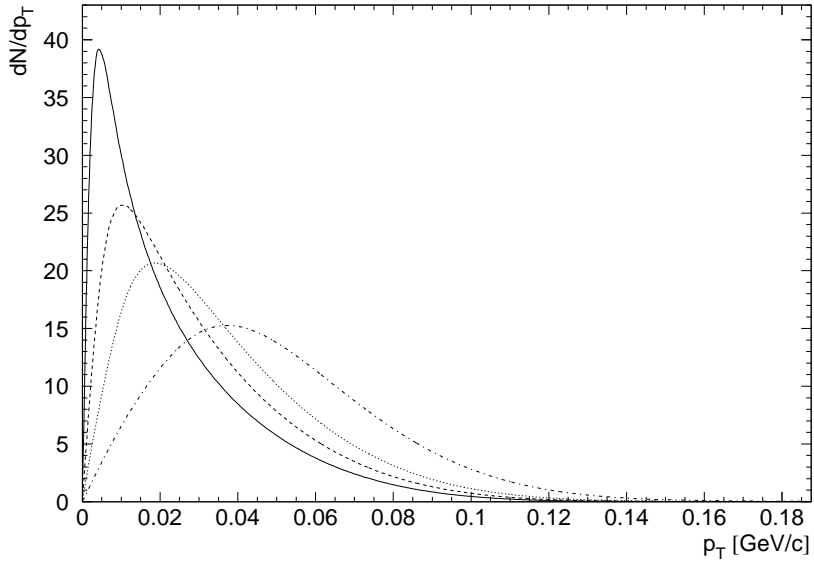}}
\caption[]{The transverse momentum spectrum $dN/dp_T$ at
mid-rapidity, $y=0$, for two-photon production of states with masses of
0.2 (solid), 0.5 (dashed), 1.0 (dotted) and 3.0 (dash-dotted) GeV in
Au + Au interactions at RHIC.  The curves are normalized to have equal
areas \cite{joakim}.}
\label{ptspectrum}
\end{figure}

Figure \ref{ptspectrum} shows the $p_T$ spectra  at 
mid-rapidity, $y=0$, for two-photon production of states with masses of
0.2, 0.5, 1.0 and 3.0 GeV in AuAu collisions at RHIC.  
The spectra peak around $\sqrt{1.5}M/\gamma$.  The average $p_T$ is around 75\% larger,
because of the long high-$p_T$ tail.  

In photonuclear interactions, most of the $p_T$ comes from the nucleus; the typical
$p_T$ is is a few $\hbar/R_A$, or 50-100 MeV/c \cite{rhotag,STAR}.  This is several times higher than is typical
for two-photon interactions.  Particularly at the LHC, where $M/\gamma$ is very small,
$p_T$ should be an effective tool for separating the two event classes.

\section{Conclusions}

We have calculated the total cross sections and rapidity and
transverse momentum distributions for two--photon production of
various final states with and without nuclear breakup.  The cross-sections
for $\gamma\gamma$ interactions accompanied by $XnXn$ and $1n1n$ dissociation
are about 1/10 and 1/100 of that for the unaccompanied $\gamma\gamma$ interactions.
The nuclear breakup tagged events have different rapidity and $p_T$ distributions
from the un-tagged events, and can be used to explore the effects of
different photon spectra and $b$ distributions. The typical $p_T$ of $\gamma\gamma$
final states is several times smaller than for comparable coherent photonuclear interactions,
so $p_T$ cuts may be an effective tool for separating the two interactions. 

This work was supported by the U.S. Department of Energy under 
Contracts No. DE-AC-03076SF00098 and DE-AC02-98CH10886.   

\end{document}